\documentclass[aps,preprint]{revtex4}
\usepackage{graphicx}

\begin{document}

\title{Entropy landscape and non-Gibbs  solutions in constraint satisfaction problems}

\author{L. Dall'Asta}
\email{dallasta@ictp.it}

\author{A. Ramezanpour}
\email{aramezan@ictp.it}

\affiliation{The Abdus Salam International Centre for Theoretical Physics, Strada Costiera 11, 34014 Trieste, Italy}

\author{R. Zecchina}
\email{riccardo.zecchina@polito.it}

 \affiliation{Politecnico di Torino, C.so Duca degli Abruzzi 24, I-10129 Torino, Italy}

\date{\today}

\begin{abstract}
We study the entropy landscape of solutions for the  bicoloring problem in random graphs,
a representative difficult constraint satisfaction problem.
Our goal is to classify  which type of clusters of solutions are addressed by  different algorithms.
In the first part of the study we use the cavity method to obtain the number of clusters
with a given internal entropy and determine the phase diagram of the problem, e.g.
dynamical, rigidity  and SAT-UNSAT transitions.
In the second part of the paper we analyze different  algorithms and locate their behavior
in the entropy landscape of the problem. For instance we show that a smoothed version of
a decimation strategy based on Belief Propagation is able to find solutions belonging to
sub-dominant clusters  even beyond   the so called rigidity transition where the thermodynamically
relevant clusters become frozen.  These non-equilibrium  solutions belong  to the most probable unfrozen clusters.
\end{abstract}

 \maketitle

\section{Introduction and Motivations}\label{0}

Many disciplines have at their root random Constraint Satisfaction Problems (CSPs).
Examples  are Information Theory where they are used to design  error correcting codes \cite{G-book-1968,M-book-2002} or
Computer Science where they constitute elementary  models for studying the onset of exponential
regimes in algorithms\cite{DMSZ-tcs-2001}. More in general, random CSPs capture some of the optimization aspects of
complex systems found in physics (e.g. spin-glasses and  packing problems),
in economics (e.g. financial markets)\cite{CMZ-prl-2000,GLMSWZ-prl-2006} and in biology (e.g. gene networks reconstruction
and  learning in neuroscience \cite{BZ-prl-2006,BBBZ-pnas-2007}).
A  random CSP is characterized by  an extensive list  of constraints,  each one
forbidding some of the joint assignments of  the (discrete) variables it involves.
In packing  problems for instance,  overlapping positions of the  elementary tiles on a given
lattice are forbidden.
Given an instance of a CSP, one wants to know whether there exists a solution,
that is an assignment of the variables which satisfies all the constraints
(e.g. a proper tiling or a proper coloring of a graph).  When such assignment  exists the instance is called SAT,
and one wants to find it. Most of the interesting CSPs  are NP-complete\cite{C-acm-1971,GJ-book-1979}: in the
worst case the number of operations needed to decide whether an instance is
SAT or not is expected to grow exponentially with the number of variables.
The interesting limit for random CSP is the thermodynamic one where both the number $N$ of
independent variables and the number $M$ of constraints go to infinity
at fixed constraint density $\alpha \equiv M/N$. The  most intriguing  phenomenon is
certainly the appearance of  sharp thresholds \cite{KS-science-1994,MZKST-nature-1999}.
At some  critical ratio $\alpha_s$  the probability of existence of solutions jumps from one to zero.
Just below such threshold, most of the known heuristic algorithms are observed to undergo a
dramatic slowing down. Such  phenomenon has been put in connection  with the onset of
a clustering phase, where the space of solution becomes divided in a large (exponential) number of different
clusters (or states) and variables develop non trivial long range correlations.

Scope of this study is to go a step further in the statistical physics analysis of the connection between  clustering
and the behavior of algorithms.  By an analytic estimate of  the internal entropy of the clusters  found  by
different algorithms on large problem instances, and by a large deviation analysis of clusters distribution with 
respect to their internal entropy, we
are able to display which type of clusters are addressed by different algorithms.
For random CSP  in the clustering phase, we observe  that local search algorithms may  be attracted by a large spectrum
of clusters  (not surprisingly as it happens in out-of-equilibrium physical systems).
Quite surprisingly we also show  that there exist simple  message-passing (MP) processes that are capable of  finding efficiently solutions
even in the harder region where the thermodynamically  dominant clusters become frozen. In such region local search algorithms are
observed numerically  to undergo an exponential slowing down  due to the global rearrangements
needed  to correct the errors made along the search process.
On the contrary, the MP  processes that we study  continue to   find solution  efficiently by addressing  clusters
which are more rare than the dominating ones (i.e. those which would be seen by sampling solutions with  uniform measure over
the solution space).
These results, together with the evidence coming from fully connected CSP that even  frozen solutions may  be found by MP \cite{BZ-prl-2006},
shed new light on how MP algorithms can be  utilized and suggest that some further  algorithmic progress may be at hand.

In what follows we first provide a very brief review of the known results and next apply our arguments to the so called Bicoloring problem,
which has some analytical advantages compared to other NP-complete problems like K-SAT or Coloring while retaining all the
conceptual features.

The paper is organized as follows. In sect. \ref{01}  an intuitive introduction to clustering is given.
The definition of the specific problem under study is provided in section \ref{1} together with a summary of previously known results.  The cavity method for large deviations is presented in section \ref{2}. The numerical methods used to solve the cavity equations and extract the complexity curves are described in section \ref{3}.
In section \ref{4} the equations are solved in some special cases in order to obtain the main properties of the phase diagram of the problem. The algorithms used to find solutions and locate them in the entropy landscape are described in section \ref{5}.
Section \ref{6} is devoted to summarize the main results of the paper and to make some concluding remarks and discuss perspectives.
In the appendices we report the details of deriving the cavity equations and 
some quantities that are essential in computing the free energy.

\section{Geometry of solutions and freezing}\label{01}

The set of solutions of a random CSP   should be thought of as  a  portion of the phase space which  may undergo a fragmentation
into   clusters for values of the density of constraints right below the SAT threshold. This scenario can be made rigorous in few
cases, the simplest one being the random XOR-SAT problem\cite{MRZ-jstatphys-2003,CDMM-prl-2003}:   the density of constraints were  clustering appears corresponds to the  percolation of particular structures in  the underlying  graph of constraints.
This fact can be used to define  clusters  which,  by linearity of XOR-SAT,  turn out to be all  identical. One may prove that  a finite fraction of the variables have to be frozen, that is  must  take the same value in all the solutions belonging to a cluster.
This picture is  however far from general:  both the definition of clusters and the analysis of their fluctuations  in size
are difficult tasks, which require the application of the cavity method in a rather  advanced setting \cite{MPR-prl-2005,M-thesis,KMRSZ-pnas-2007,ZK-arxiv-2007,S-arxiv-2007}.
One important feature of clusters in  CSPs concerns the presence or absence of frozen  variables.  It may happen that clusters with frozen variables coexist with totally unfrozen clusters of  larger internal entropy, with big effects on the hardness of the associated combinatorial optimization problem. The intuitive reason why the presence of frozen variables is believed to be relevant is well summarized by  the
idea of rearrangements \cite{S-arxiv-2007}:  "given an initial solution of a CSP and a variable  $i$ that  one would like to modify, a rearrangement is a path in  configuration space that starts from the initial solution and  leads to another solution where the value of the $i$-th variable is changed with  respect to the initial one. The minimal  length of such a path is a measure of how constrained was the variable $i$ in the initial configuration.  In intuitive terms  this length diverges with the system size when the variable was frozen in the initial cluster".
The idea is that when freezing takes place in Gibbs states, then the rearrangements are responsible for  a critical slowing down of local
search algorithms. On the contrary, when dominant clusters are not frozen, even relatively simple local algorithms may find solutions by incrementally adding constraints until the full problem is satisfied. Recent arguments and numerical studies,  have shown that one can still obtain a solution beyond  the dynamical transition as long as the so-called {\em jamming transition} has not occurred \cite{KK-arxiv-2007}. Following Ref.~\cite{KK-arxiv-2007}, one can imagine adding the constraints one by one, in each step recording the number of flips that are required in order to find the new solution. Close to the jamming transition the number of flips diverges and makes it difficult to find a solution to the CSP \cite{S-arxiv-2007}.

A Constrained Satisfaction Problem is defined by a set of constrains $\mathcal{C}=\{I_a[\underline{\sigma}]\in \{0,1\}|a=1,\ldots,M\}$ on a number of variables. The constraints depend on the configuration of variables $\underline{\sigma}\equiv \{\sigma_i|i=1,\cdots,N\}$ and the problem is called {\em satisfiable} if all constraints are satisfied, i.e. $I_a=1, \forall a$.  A {\em solution} of the problem is a configuration of the variables that satisfies all constraints. In analogy with statistical physics models, we define the {\em energy} $E[\underline{\sigma}]$ of configuration $\underline{\sigma}$ as the number of unsatisfied constraints in $\underline{\sigma}$.
Given an instance of a CSP, one is interested in deciding whether it is satisfiable (i.e. $E[\underline{\sigma}]=0$) and, in such a case, in explicitly finding solutions to the problem.

More in general, one can define an ensemble of instances of the problem considering all possible random assignments of the $M$ constraints among $N$ variables, with fixed density of constraints $\alpha = M/N$. Varying $\alpha$ it was shown that the system passes from a phase in which it is always possible to find a solution, the {\em SAT phase}, to the {\em UNSAT phase} where a fraction of the constraints are not satisfied.
Examples of studies for root problems such as Random Satisfiability Problem, XOR-SAT and Graph Coloring can be found in \cite{MZ-pre-1997,MPZ-science-2002,MZ-pre-2002,MPWZ-prl-2002,ANP-nature-2005,MMZ-rsa-2006}.
The main  tool for  analyzing the satisfiability of typical problem  instances is the {\em cavity method}, originally developed to study the thermodynamic properties of diluted spin-glass systems \cite{MPV-1987} and recently reconsidered in the context of CSPs \cite{MP-epjb-2001,MP-jstatphys-2003,MZ-pre-2002}.  Actually, the cavity method at the ensemble level allows to study the typical properties as well as large deviations from typical behaviors \cite{R-jstatmech-2005,MPR-prl-2005,KMRSZ-pnas-2007}.
The key feature of the cavity method which is of interest for computer science stems from the discovery that it can be converted to an algorithm for analyzing  single problem instances \cite{MZ-pre-2002,BMZ-rsalgorithm-2005}, becoming an efficient  solver for the random  combinatorial problems.

\section{Definition of the problem and known results}\label{1}

\begin{figure}
\includegraphics[width=10cm]{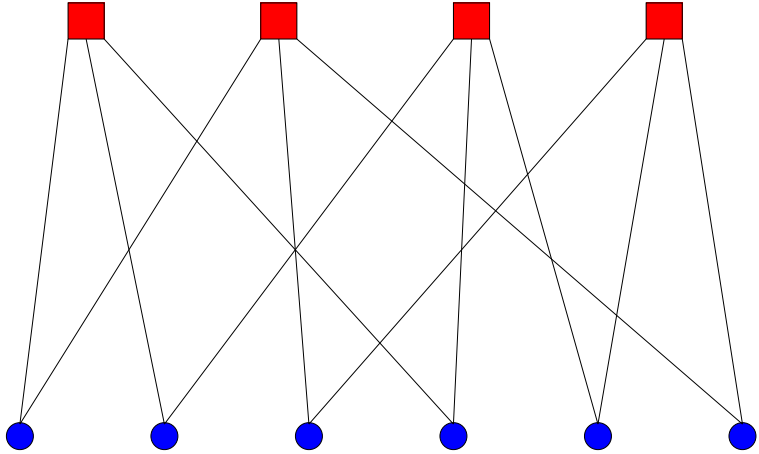}
\caption{A regular random factor graph with function nodes (squares) of degree $K=3$ and variable nodes (circles) of degree $L=2$.}\label{figure1}
\end{figure}

Consider a hypergraph of $N$ nodes $i=1,\ldots,N$ and $M$ hyperedges $a=1,\ldots,M$. For simplicity we consider regular random hypergraphs, or $(K,L)$-hypergraphs, where each hyperedge connects $K$ nodes and each node contributes in $L=KM/N=K\alpha$ hyperedges. A node in this hypergraph has state $\sigma_i \in \{0,1\}$ and each hyperedge imposes a constraint on the values of the associated variables. In the bicoloring problem this constraint just forbids the configurations in which  all the nodes belonging to an hyperedge have the same value. In the context of circuit logic the bicoloring problem is known as Not-All-Equal-Satisfiability (NAE-SAT) problem, in physics it is a spin model with anti-ferromagnetic interactions.

We may represent bicoloring as a factor graph \cite{KFL-ieee-2001}. This is a bipartite graph where  the variables and constraints are represented with two different types of nodes, variable and function nodes, respectively. Each function node is connected to all the variable nodes that should satisfy the associated constraint. Figure \ref{figure1} shows an example of regular random factor graph.

The hypergraph bicoloring problem is NP-complete for $K\ge 3$ \cite{F-jam-1999}.
The case $K=3$ with a Poisson degree distribution for the variable nodes has already been studied in \cite{CNRZ-jphysa-2003}.
The authors found dynamical and SAT/UNSAT transitions within the single and multiple cluster approximations. In spin glass language these approximations are called replica symmetric (RS) and one-step replica symmetry breaking (1RSB) approximation, respectively. In the latter case the authors only consider the most numerous  clusters.

\begin{table}

\begin{center}

\begin{tabular}{|c|c|c|c|}
  \hline

    $K$      &  $L^0_d$ & $L_s^{RS}$ & $L_s^{1RSB}$   \\
  \hline

   $3$    & $-$  & $8$ & $7$  \\

  \hline

   $4$   & $17$  & $21$ & $20$  \\

  \hline

   $5$   & $40$  & $54$ & $53$  \\

  \hline

   $6$    & $91$ & $131$ & $130$  \\

  \hline

   $7$    & $200$ & $309$ & $307$  \\

  \hline

   $8$    & $428$ & $708$ & $705$  \\

  \hline

   $9$   & $905$  & $1594$  & $1592$\\

  \hline

   $10$   & $1894$  & $3546$ & $3543$ \\

  \hline

\end{tabular}

\vskip 0.5cm

\caption{Numerical values of $L^0_d$ and $L_s$ (in the RS and 1RSB approximations).  In each case we have given the smallest integer degree larger than or equal to the precise value.}\label{table1}
\end{center}
\end{table}

The intensive entropy $s$ is defined by the number of solutions $\mathcal{N}=e^{Ns}$. Using Bethe approximation in the replica symmetric phase we find the entropy as

\begin{equation}\label{sRS}
s^{RS}=\ln [2(1-\frac{1}{2^{K-1}})^L]-(K-1)\alpha \ln [1-\frac{2}{2^K}].
\end{equation}
This quantity vanishes at
\begin{equation}
L_s^{RS}=-K\frac{\ln2}{\ln(1-\frac{1}{2^{K-1}})},
\end{equation}
where for $K\gg 1$ gives
\begin{equation}\label{LsRS}
L_s^{RS} \simeq K 2^{K-1} \ln2(1+O(\frac{1}{2^K})) .
\end{equation}
If there exist more than one cluster of solutions we define the complexity $\Sigma$ by $\mathcal{N}_c=e^{N\Sigma}$ where $\mathcal{N}_c$ is the number of clusters. Notice that for very large $N$ the above complexity is dominated by typical clusters.  In 1RSB approximation and considering only typical clusters, the complexity reads \cite{CNRZ-jphysa-2003},
\begin{equation}
\Sigma_{typ}=\ln[A_L]-(K-1)\alpha \ln[1-\frac{1}{2}(1-\eta)(1-\frac{\eta^{L-1}}{A_{L-1}})],
\end{equation}
where
\begin{equation}\label{AL}
A_L=2(1-\frac{1-\eta}{2})^L-\eta^L,
\end{equation}
and $\eta$ is determined by the following equation
\begin{equation}\label{eta}
\eta=1-2[\frac{1}{2}(1-\frac{\eta^{L-1}}{A_{L-1}})]^{K-1}.
\end{equation}
A nontrivial solution ($\eta \ne 1$) for the above equation results to a nonzero complexity. 
Let us assume $K\gg 1$ and find the point where for the first time a nontrivial solution appears.
We try $\eta=1-\frac{c}{2^{K-1}}$ and find $c$ in a self-consistent way.
From the above equation we obtain
\begin{equation}
\frac{c}{2}\simeq  \exp[-\frac{K-1}{2}e^{-c\frac{L-1}{2^{K}}}].
\end{equation}
It means that to have a finite solution for $c$ we need $L$ diverges as
\begin{equation}\label{L0d}
L^0_d \simeq 2^{K}\frac{1}{c}[\ln K-\ln 2-\ln \ln(\frac{2}{c})+o(1)].
\end{equation}
At the SAT/UNSAT transition $\Sigma_{typ}$ vanishes and we can use this fact to determine $\alpha_s$. This value behaves, asymptotically, as $\alpha_s^{RS}$ in Eq. \ref{LsRS}. In table \ref{table1} we compare the numerical values of $L_d^0$ and $L_s$ obtained with the above methods.

\section{Cavity method: a large deviation study}\label{2}

A more complete picture of the distribution of clusters is given by a large deviation study \cite{R-jstatmech-2005,MPR-prl-2005,KMRSZ-pnas-2007}. We define the partition function at zero temperature as
\begin{equation}\label{Z}
Z=\sum_{\underline{\sigma}}\prod_a I_a(\sigma_{\partial a
})e^{x\sum_i (\sigma_i-\sigma_i^*)^2}.
\end{equation}
Here $x$ is a Lagrange multiplier that controls the distance between the solutions and a reference point $\underline{\sigma}^*$. And $\sigma_{\partial a}=\{\sigma_i|i\in V(a)\}$ where $V(a)$ is the set of neighbors of function node $a$. For $x=0$ we recover the total number of solutions.
If there is only one cluster of solutions we can safely use the standard {\em Belief Propagation} (BP) equations \cite{KFL-ieee-2001} to obtain an estimate of the cavity marginals (see appendix A)
\begin{equation}\label{muia}
\mu_{i\rightarrow a}(\sigma_i)=\frac{1}{Z_{i\rightarrow
a}}\sum_{\sigma_{\partial i\rightarrow a}}\left(\prod_{b\in V(i)\setminus a}
I_b(\sigma_{\partial b })[\prod_{j\in V(b)\setminus i} \mu_{j\rightarrow
b}(\sigma_j)]\right)e^{x(\sigma_i-\sigma_i^*)^2}.
\end{equation}
Here $Z_{i\rightarrow a}$ is a normalization constant, $V(i)$ is the set of neighbors of variable node $i$ and $\sigma_{\partial i\rightarrow a}=\{ \sigma_j|j\in V(b), b\in V(i)\setminus a\}$. We will write the above equation in short as
\begin{equation}\label{Smu}
\mu_{i\rightarrow a}(\sigma_i)=\mathcal{S}[\mu_{j\rightarrow b}].
\end{equation}
We also define the free energy $f(x)$ as
\begin{equation}
Z=e^{Nf(x)}=\sum_d e^{N[s(d)+xd]},
\end{equation}
where $d=\frac{1}{N} \sum_i(\sigma_i-\sigma_i^*)^2$ and $e^{Ns(d)}$ is the number of solutions at distance $d$ from the reference point. In the Bethe approximation
\begin{eqnarray}\label{free_energy1}
f(x)=\sum_i \Delta f_i-\sum_a
(K_a-1)\Delta f_a,
\end{eqnarray}
where
\begin{eqnarray}\label{dfidfa}
e^{N\Delta f_i}= \sum_{\sigma_i}[\prod_{a\in V(i)}
\mu_{a\rightarrow
i}(\sigma_i)]e^{x(\sigma_i-\sigma_i^*)^2},\\ \nonumber
e^{N\Delta f_a}= \sum_{\sigma_{\partial a}}I_a(\sigma_{\partial
a})\prod_{i\in V(a)} \mu_{i\rightarrow a}(\sigma_i).
\end{eqnarray}

Using the above free energy we can determine the entropy $s(d)$ by a Legendre transform
\begin{equation}\label{entropy1}
s(d)=\min_{x}{[f(x)-xd(x)]}, \hskip1cm d(x)=\frac{\partial f(x)}{\partial x}.
\end{equation}

If the replica symmetry is broken, we would have different clusters of solutions and the cavity marginals fluctuate from one cluster to another one. This is described by  a generalized partition function defined by
\begin{equation}\label{mathcalZ}
\mathcal{Z}\equiv e^{N\mathcal{F}(m)}=\sum_{c} e^{m Ns_c}= \int ds
e^{N[\Sigma(s)+ms]}.
\end{equation}
Here $c$ labels the clusters and $s_c$ is the internal entropy of cluster $c$. Again in the Bethe approximation
\begin{eqnarray}\label{free_energy2}
N\mathcal{F}(m)=\sum_i \ln \mathcal{Z}_i-\sum_a (K_a-1)\ln \mathcal{Z}_a
\end{eqnarray}
where
\begin{eqnarray}\label{mathcalZiZa}
\mathcal{Z}_i\equiv \int \prod_{a\in V(i)}\prod_{j\in
V(a)\setminus i} d\mathcal{P}_{j\rightarrow a}[\mu_{j\rightarrow a}]
e^{mN\Delta s_i}, \\ \nonumber
\mathcal{Z}_a\equiv \int \prod_{i \in
V(a)}d\mathcal{P}_{i\rightarrow a}[\mu_{i\rightarrow a}]
e^{mN\Delta s_a}.
\end{eqnarray}

Having the generalized free energy we can determine the complexity $\Sigma(s)$ by a Legendre transformation
\begin{equation}\label{complexity}
\Sigma(s)=\min_m{[\mathcal{F}(m)-ms]}, \hskip1cm s(m)=\frac{\partial \mathcal{F}(m)}{\partial m}.
\end{equation}
In appendix B we have explained the origin of the above relations. 
Notice that $\Delta s_i=\Delta f_i(x=0)$ and $\Delta s_a=\Delta f_a(x=0)$, where the free energy shifts are given by Eq. \ref{dfidfa}.
Moreover, $\mathcal{P}_{i\rightarrow a}[\mu_{i\rightarrow a}]$ is the probability that, in a randomly selected cluster, we find the cavity marginal $\mu_{i\rightarrow a}$ on edge $(i,a)$ of the factor graph. 
This probability distribution is determined by the following self-consistency equation:
\begin{equation}\label{pmu}
\mathcal{P}_{i\rightarrow a}[\mu_{i\rightarrow
a}]=\frac{1}{\mathcal{Z}_{i\rightarrow a}}\int \prod_{b\in
V(i)\setminus a}\prod_{j\in V(b)\setminus i} d\mathcal{P}_{j\rightarrow
b}[\mu] e^{mN\Delta s_i}\delta(\mu_{i\rightarrow
a}-\mathcal{S}[\mu]),
\end{equation}
where $\mathcal{Z}_{i\rightarrow a}$ is a normalization constant and $\mathcal{S}[\mu]$ is the same as in Eq. \ref{Smu} with $x=0$.  The factor $e^{mN\Delta s_i}$ is to sample correctly the clusters when we add the new variable $i$.
Let us multiply the two sides of Eq. \ref{pmu}  by  $2\mu_{i\rightarrow a}(\sigma)$, to find the new probability distribution $\mathcal{Q}_{i\rightarrow a}^{\sigma}[\mu]=2\mu_{i\rightarrow a}(\sigma)\mathcal{P}_{i\rightarrow a}[\mu]$ that will be useful in the special case of $m=1$. In the right hand side we can replace $\mu_{i\rightarrow a}(\sigma)$ with its definition in Eq. \ref{muia}. Rearranging the terms we get
\begin{eqnarray}\label{qmu}
\mathcal{Q}_{i\rightarrow a}^{\sigma_i}[\mu]=\frac{2}{\mathcal{Z}_{i\rightarrow a}}\int \sum_{\sigma_{\partial i\rightarrow a}}\prod_{b\in
V(i)\setminus a}I_b(\sigma_{\partial b})\prod_{j\in V(b)\setminus i} \left( \frac{1}{2}d\mathcal{Q}_{j\rightarrow
b}^{\sigma_j}[\mu]\right)  e^{(m-1)N\Delta s_i}\delta(\mu_{i\rightarrow
a}-\mathcal{S}[\mu]).
\end{eqnarray}
In the following we will split $\mathcal{P}_{i\rightarrow a}[\mu]$ into frozen and unfrozen parts as
\begin{equation}
\mathcal{P}_{i\rightarrow a}[\mu_{i\rightarrow a}]=\frac{1-\pi}{2}[\delta(r)+\delta(r-1)]+\pi \rho(r),
\end{equation}
where $\mu_{i\rightarrow a}(0)=r$, $\mu_{i\rightarrow a}(1)=1-r$ and $\rho(r)$ is the probability distribution of unfrozen fields.
The above arguments become much simpler in random $(K,L)$-hypergraphs where all the links and nodes are statistically equivalent. For example, Eq. \ref{free_energy2} is replaced with
\begin{equation}
\mathcal{F}(m)=\ln[\int D_i\mathcal{P}[\mu] e^{mN\Delta
s_i}]-\alpha (K-1)\ln[\int D_a\mathcal{P}[\mu] e^{mN\Delta s_a}],
\end{equation}
where $\int D_i\mathcal{P}[\mu]$ and $\int D_a\mathcal{P}[\mu]$ denote the integrations over all the cavity marginals that contribute in $\Delta s_i$ and $\Delta s_a$, respectively.

\section{Entropy landscape: Numerical method}\label{3}

The main equation that we should solve is Eq.\ref{pmu}. One can use the population dynamics method \cite{MP-epjb-2001,MP-jstatphys-2003} to get rid of summing over a large number of continuous variables.

\subsection{In a single hypergraph}\label{31}

\begin{figure}
\includegraphics[width=10cm]{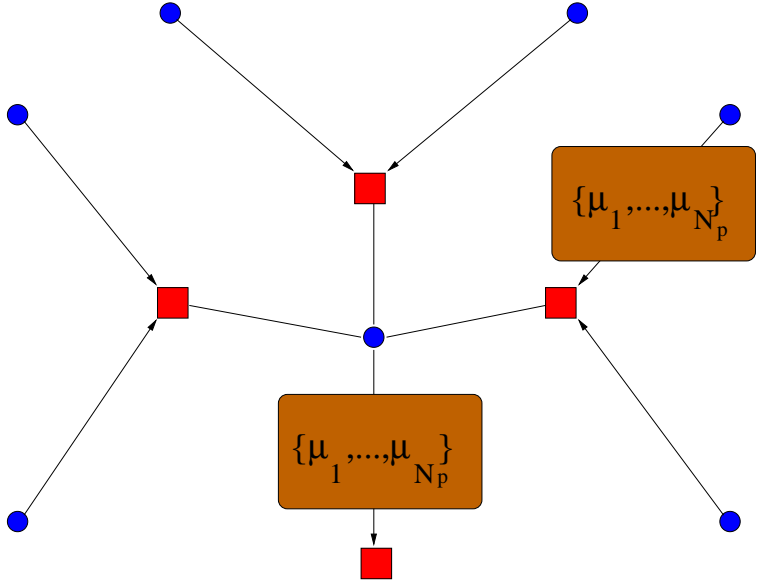}
\caption{Population dynamics works with a population of fields on each link of the factor graph.}\label{figure2}
\end{figure}

Given the factor graph we represent $\mathcal{P}_{i\rightarrow a}[\mu]$ by a population of $\mathcal{N}_p$ cavity probabilities (or fields), Fig. \ref{figure2}. At the initial point all the cavity fields are of frozen kind, with equal probability for $r=0$ and $r=1$.
With this initial condition we will not miss a nontrivial solution with frozen fields, if any.
At each step of the population dynamics we select a link $(i\rightarrow a)$ randomly and do in the following way:

\begin{itemize}

\item For each $b\in V(i)\setminus a$ and $j\in V(b)\setminus i$: randomly select a member of the population on link $(j\rightarrow b)$.

\item Using these $(L-1)(K-1)$ fields: calculate the new $\mu_{i\rightarrow a}$ by Eq. \ref{muia}.

\item Calculate the weight $w_{i\rightarrow a}=e^{mN\Delta s_i}$ from Eq. \ref{dfidfa} at $x=0$. This weight is zero if there is any contradiction.

\item With probability $\frac{w_{i\rightarrow a}}{w^{max}_{i\rightarrow a}}$ replace a randomly selected member of the population with the new one. Here $w^{max}_{i\rightarrow a}$ is the maximum weight $w_{i\rightarrow a}$ observed in the evolution from the beginning.

\end{itemize}

\begin{figure}
\includegraphics[width=10cm]{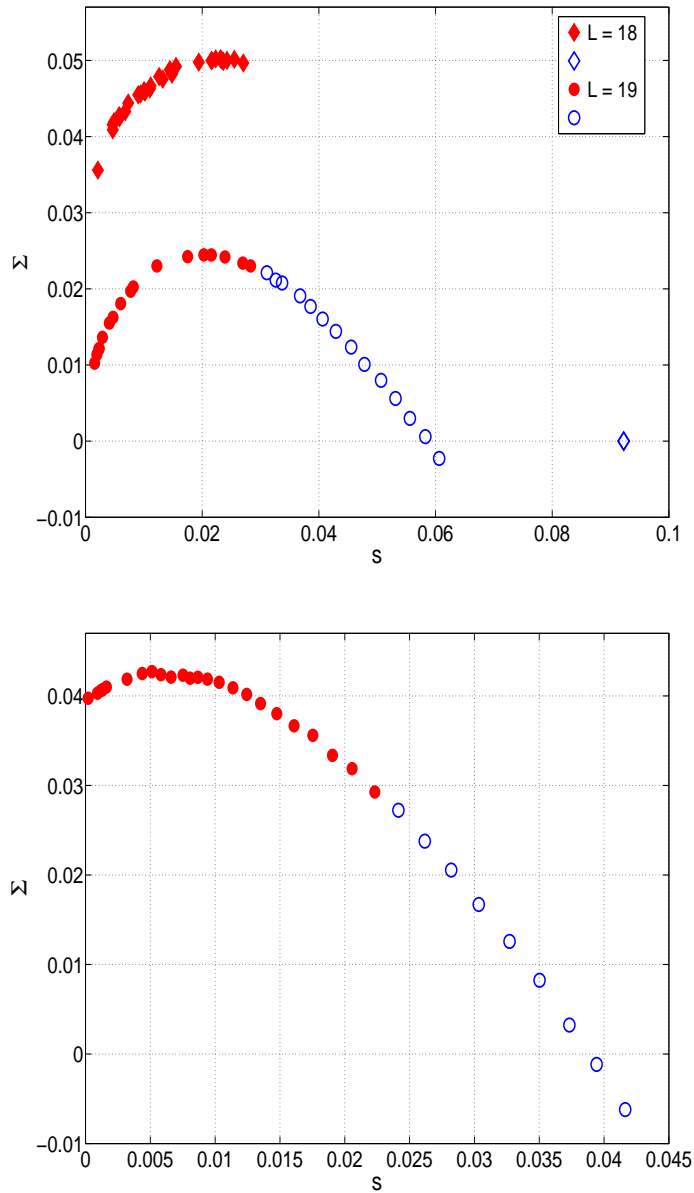} 
\caption{Complexity vs entropy for $K=4$, $L=18,19$ with $\mathcal{N}_p=10^5$ (top) and $K=6$, $L=121$ with $\mathcal{N}_p=2\times 10^4$ (bottom) in one-link approximation . Filled and empty symbols
represent frozen and unfrozen parts, respectively. The statistical errors are about $0.001$.}\label{figure3}
\end{figure}

In a sweep of the algorithm we choose all the links of the factor graph  randomly.  Having the populations we can obtain the free energy as
\begin{eqnarray}\label{free_energypop}
N\mathcal{F}(m)=\sum_i \ln \mathcal{Z}_i-(K-1)\sum_a \ln \mathcal{Z}_a, \\ \nonumber
\mathcal{Z}_i=\langle e^{mN\Delta
s_i}\rangle_{pop},\\ \nonumber
\mathcal{Z}_a=\langle e^{mN\Delta s_a} \rangle_{pop},
\end{eqnarray}
where $\langle. \rangle_{pop}$ means averaging over the populations. We stop the updates as soon as the free energy, and so the weights $w^{max}_{i\rightarrow a}$,  reach the steady state. Then the entropy reads
\begin{eqnarray}\label{entropypop}
Ns(m)=\sum_i \overline{\Delta s_i}-(K-1)\sum_a \overline{\Delta s_a}, \\ \nonumber
\overline{\Delta s_i}=\frac{\langle \Delta s_i e^{mN\Delta
s_i}\rangle_{pop}}{\langle e^{mN\Delta
s_i}\rangle_{pop}} ,\\ \nonumber
\overline{\Delta s_a}=\frac{\langle \Delta s_a e^{mN\Delta
s_a}\rangle_{pop}}{\langle e^{mN\Delta
s_a}\rangle_{pop}}.
\end{eqnarray}

Figure \ref{figure3}  shows the results for choices of the factor graph parameters that correspond to different phases.

\subsection{In one-link approximation}\label{32}

\begin{figure}
\includegraphics[width=15cm]{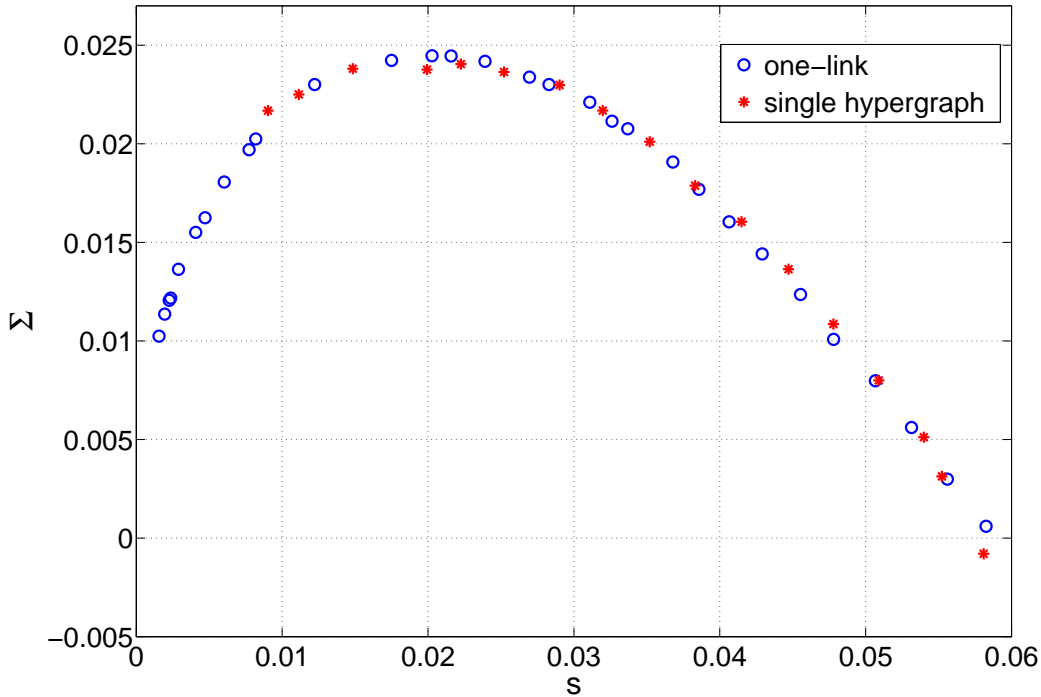}
\caption{Comparing $\Sigma(s)$ in a single $(4,19)$-hypergraph ($N=10^4$, $\mathcal{N}_p=10^2$) and in one-link approximation with $\mathcal{N}_p=10^5$.  The statistical errors are about $0.001$.}\label{figure4}
\end{figure}

In a regular random hypergraph all links of the associated factor graph are equivalent. Therefore we can forget about different populations on different links and work with only one large population of fields.
The way we obtain the stationary distribution $\mathcal{P}[\mu]$ is the same as above. The only difference is that we always select the fields from the single population.  In Fig. \ref{figure4} we compare the complexity computed on a single $(4,19)$-hypergraph of size $N=10^4$ with the complexity obtained in the one-link approximation.

\section{Entropy landscape: Analytical results}\label{4}

To locate different phase transitions in the solution space
we need to calculate the generalized free energy $\mathcal{F}$ which is given in terms of $\mathcal{Z}_i$ and $\mathcal{Z}_a$ in Eq. \ref{mathcalZiZa}. These quantities in turn depend on the fraction of frozen variables $\pi$ and $\rho(r)$ which should be determined by Eq. \ref{pmu} for $\mathcal{P}[\mu]$.
In the following we study some special cases that allow us to calculate the above quantities and determine the phase diagram of the problem. For clarity here we only state the results of calculations that will be presented in more details in appendix C.

\subsection{The case $m=1$}\label{41}

The study of $m=1$ clusters is relevant in determining the dynamical,  rigidity and condensation transitions \cite{KMRSZ-pnas-2007}. These are in fact the thermodynamically relevant clusters before the condensation transition. 

For $m=1$ the generalized free energy reads
\begin{equation}
\mathcal{F}=\ln\left(2[1-\frac{1}{2^{K-1}}]^{L} \right)-(K-1)\alpha \ln\left(1-\frac{2}{2^K}\right).
\end{equation}
Comparing with the RS entropy $s^{RS}$ in Eq. \ref{sRS} we see that $\mathcal{F}(m=1)=\Sigma(m=1)+s(m=1)=s^{RS}$. Therefore, as long as the $m=1$ clusters are the thermodynamically relevant ones the RS approximation gives the correct total entropy.
From  Eq. \ref{qmu} we can also find the probability of having a frozen field with $r=1$
\begin{equation}
\frac{1-\pi}{2}=\frac{1}{\mathcal{Z}_{i\rightarrow a}}\left([1-\frac{1}{2^{K-1}}]^{L-1}-[1-\frac{1}{2^{K-1}}-(\frac{1-\pi}{2})^{K-1}]^{L-1}\right).
\end{equation}
For small $L$ the above equation has only one solution, $\pi=1$, where the $m=1$ clusters are unfrozen. Increasing $L$, one reaches the rigidity point $L_r$ where another solution $\pi\ne 1$ appears. It is where a finite fraction of the variables in these clusters become frozen.
We find that for $K< 6$ the rigidity transition always happens after the SAT/UNSAT transition.
Simplifying the above equation we obtain
\begin{equation}
\pi=[1-\frac{(\frac{1-\pi}{2})^{K-1}}{1-\frac{1}{2^{K-1}}}]^{L-1}.
\end{equation}
Assuming $K\gg 1$ and $\pi=\frac{c}{K}$  one obtains
\begin{equation}
c \simeq K \exp[-\frac{(L-1)}{2^{K-1}}e^{-c}],
\end{equation}
which suggests
\begin{equation}
L_r \simeq 2^{K-1} e^{c}[\ln K-\ln c+o(1)].
\end{equation}
We see that, like $L_d^0$, the leading term in $L_r$ diverges as $2^K\ln K$. Compare it with the leading term of $L_s^{RS}$ which scales as $2^K K$.

\subsection{The case $m=0$}\label{42}
 
The typical or most numerous clusters are the $m=0$ ones. The study of these clusters provides us with an estimate of the  SAT/UNSAT transition (in that they are the last clusters to disappear).  Indeed the previous studies of the complexity in 1RSB phase focus on these type of clusters. For $m=0$ the generalized free energy reads
\begin{equation}
\mathcal{F}=\ln\left( 2[1-(\frac{1-\pi}{2})^{K-1}]^L-[1-2(\frac{1-\pi}{2})^{K-1}]^L \right)-(K-1)\alpha \ln\left(1-2(\frac{1-\pi}{2})^K \right).
\end{equation}
Using Eq. \ref{pmu} one can easily write an equation for the fraction of frozen marginals
\begin{equation}
\frac{1-\pi}{2}=1-\frac{[1-(\frac{1-\pi}{2})^{K-1}]^{L-1}}{2[1-(\frac{1-\pi}{2})^{K-1}]^{L-1}-[1-2(\frac{1-\pi}{2})^{K-1}]^{L-1}}.
\end{equation}
The above equation is another way of writing  Eq. \ref{eta} that has been obtained in the previous studies. Notice that
\begin{equation}
\pi=\frac{\eta^{L-1}}{A_{L-1}},
\end{equation}
with $A_{L-1}$ and $\eta$ given by Eqs. \ref{AL} and \ref{eta}.
A nontrivial solution for $\pi$ appears at $L^0_d$, where for the first time a maximum appears in the curve $\Sigma(s)$.
According to Eq. \ref{complexity} the complexity of $m=0$ clusters is $\Sigma(m=0)=\mathcal{F}$. The point that this quantity vanishes defines the SAT/UNSAT transition $L_s$. One can show that, like $L_s^{RS}$, the leading term in $L_s$ scales as $2^K K$.

\subsection{The case $\pi=0$}\label{43}

\begin{figure}
\includegraphics[width=15cm]{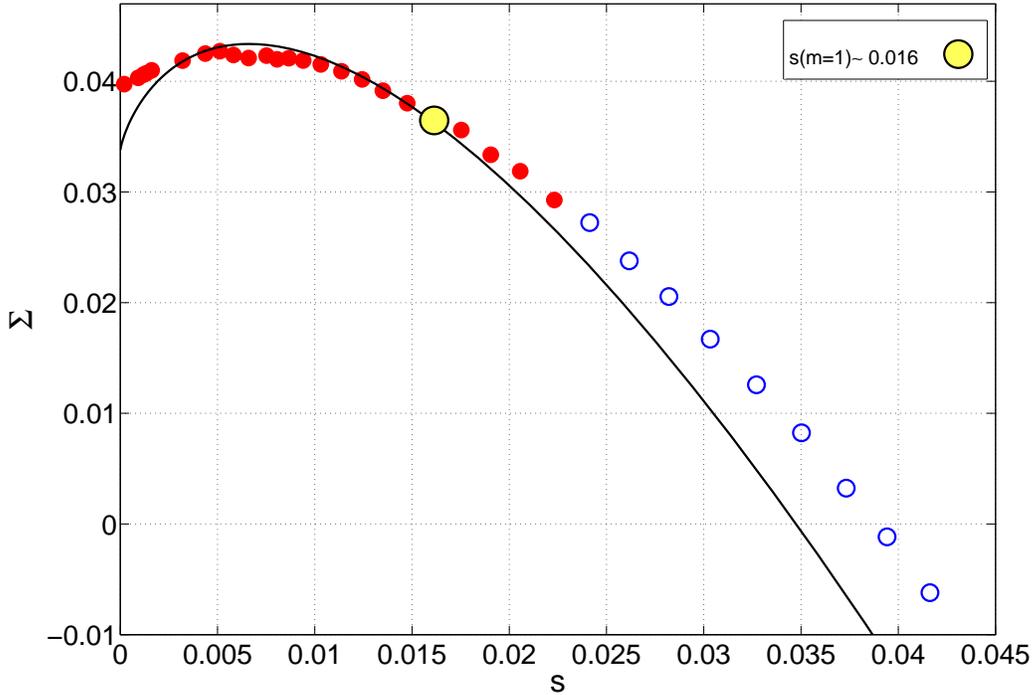}
\caption{Comparing $\Sigma(s)$ for $K=6$ and $L=121$ (in one-link approximation with $\mathcal{N}_p=2\times10^4$) with $\Sigma(s,\pi=0)$.}\label{figure5}
\end{figure}

This case is relevant to study very small clusters or close to the SAT/UNSAT transition where almost all variables are frozen and $\pi \simeq 0$.  Notice that solving numerically for $\Sigma(s)$ is a heavy computational job and it would be useful to have other approximation methods to get a good estimate of the complexity. When $\pi=0$ the generalized free energy is given by
\begin{eqnarray}
\mathcal{F}=\ln \left(2(2^{m-1}-1)[1-\frac{2}{2^{K-1}}]^L+2[1-\frac{1}{2^{K-1}}]^L \right) -(K-1)\alpha \ln \left(1-\frac{2}{2^{K}} \right).
\end{eqnarray}
Taking derivatives we obtain the entropy as
\begin{eqnarray}
s=\frac{2^m [1-\frac{2}{2^{K-1}}]^L \ln 2}{2(2^{m-1}-1)[1-\frac{2}{2^{K-1}}]^L+2[1-\frac{1}{2^{K-1}}]^L}.
\end{eqnarray}
With the above quantities we can obtain the complexity of different clusters. In Fig. \ref{figure5} we have compared this complexity with the one obtained numerically in the one-link approximation. As the figure shows the agreement is good especially for the smaller and frozen clusters.

Close to the condensation transition the $m=1$ clusters are nearly completely frozen and we expect the $\pi=0$ complexity to give a good estimate of $\Sigma(m=1)$. From the above equations we obtain
\begin{eqnarray}
\Sigma(m=1)\simeq \ln \left( 2[1-\frac{1}{2^{K-1}}]^L\right)-(K-1)\alpha \ln\left(1-\frac{2}{2^{K}} \right)
-\left(\frac{1-\frac{2}{2^{K-1}} }{1-\frac{1}{2^{K-1}}}\right)^L \ln 2.
\end{eqnarray}
We use this approximated complexity  to  determine the condensation transition, $L_c$, where $\Sigma(m=1)$ vanishes.
After some algebra we find that for $K\gg 1$
\begin{eqnarray}
L_c\simeq K2^{K-1}(\ln 2-\frac{K}{2^K}+o(1)).
\end{eqnarray}
Notice that the leading term in $L_c$ is exactly the same as in $L_s^{RS}$. 

\subsection{The case $\pi=1$}\label{44}

The complexity can be nonzero even when the frozen fields are absent. In this case we can exactly compute the free energies of $m=0,1,2$ clusters. We use this fact to find an approximated free energy for $m\in [0,2]$. 
Having $\mathcal{F}(m=0)$, $\mathcal{F}(m=1)$ and $\mathcal{F}(m=2)$ we use the Lagrange interpolating polynomial function to write $\mathcal{F}(m)$  around $m=1$ 
\begin{eqnarray}
\mathcal{F}(m)=-\mathcal{F}(m=1)m(m-2)+\mathcal{F}(m=2)\frac{m(m-1)}{2},
\end{eqnarray}
where we have used the fact that for $\pi=1$, $\mathcal{F}(m=0)=0$. The resulted entropy and complexity are
\begin{eqnarray}
s(m)=-2(m-1)\mathcal{F}+(m-\frac{1}{2})\mathcal{F}(m=2)\\ \nonumber
\Sigma(m)=m^2[\mathcal{F}(m=1)-\frac{1}{2}\mathcal{F}(m=2)].
\end{eqnarray}
As we show in appendix C, the free energy $\mathcal{F}(m=2)$ depends on the second moment of the $\rho(r)$,
\begin{eqnarray}\label{m2r2}
\langle r^2\rangle=\frac{1}{\mathcal{Z}_{i\rightarrow a}(m=2)}[1-\frac{2}{2^{K-1}}+\langle r^2\rangle^{K-1}]^{L-1}.
\end{eqnarray}
It turns out that $\Sigma(m=1)$ is zero as long as $\langle r^2\rangle=1/4$, i.e. $\rho(r)=\delta(r-\frac{1}{2})$. The complexity becomes nonzero only when equation \ref{m2r2} suggests a nontrivial solution. We can rewrite Eq. \ref{m2r2} as
\begin{eqnarray}
x=[1-\frac{1-x^{K-1}}{1+(1+x)^{K-1}(2^{K-1}-2)}]^{L-1}, x\equiv \frac{1}{2\langle r^2\rangle}-1.
\end{eqnarray}
Taking $K\gg 1$ and $x=\frac{c}{K}$ we find
\begin{eqnarray}
c\simeq K \exp[-\frac{L}{2^{K-1}}e^{-c}].
\end{eqnarray}
The equation suggests that
\begin{eqnarray}
L_d\simeq 2^{K-1}e^{c}[\ln K-\ln c+o(1)],
\end{eqnarray}
which behaves very similar to $L_d^0$ in Eq. \ref{L0d}.

\subsection{The case of integer $m$}\label{45}

\begin{figure}
\includegraphics[width=15cm]{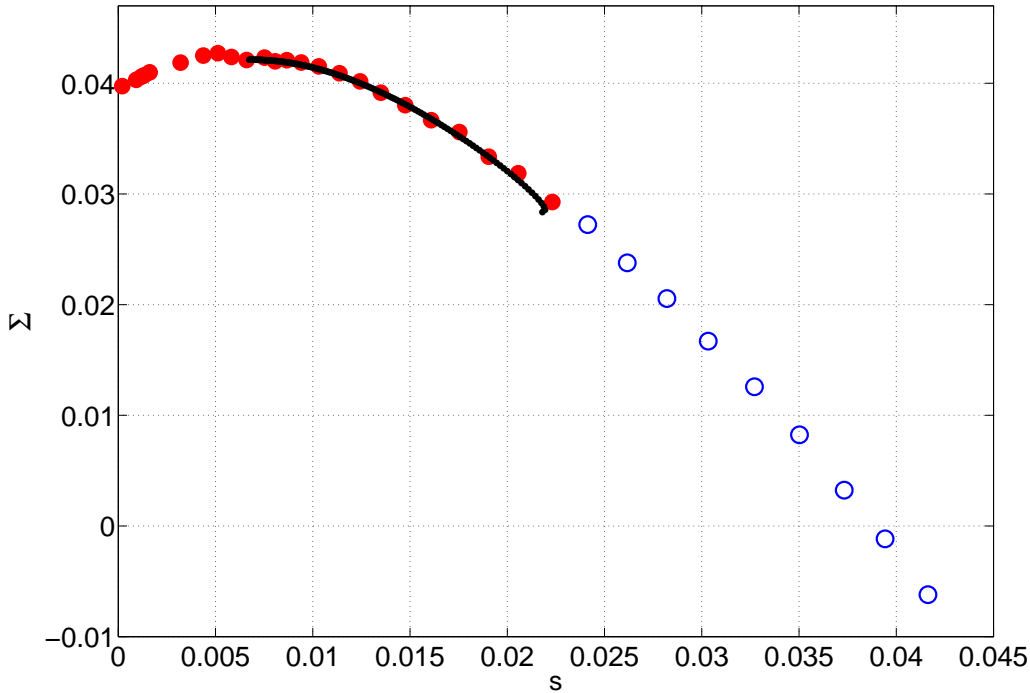}
\caption{Comparing $\Sigma(s)$ for $K=6$ and $L=121$ (in one-link approximation with $\mathcal{N}_p=2\times10^4$) with the
complexity that has been obtained by interpolation approximation ($N_m=10$).}\label{figure6}
\end{figure}

Suppose that we have computed  $\mathcal{Z}_i(m_n)$ and $\mathcal{Z}_a(m_n)$ for $m_n=0,1,\ldots,N_m-1$.  In appendix C we write explicit relations for these quantities when $\rho(r)=\delta(r-\frac{1}{2})$. We can find an approximated free energy that interpolates between the free energy values at integer $m$'s, $\mathcal{F}(m_n)$. To this end we use the Lagrange interpolating polynomial
\begin{eqnarray}\label{fenergyinter}
\mathcal{F}(m)=\sum_{n=0}^{N_m-1} \mathcal{F}(m_n)\prod_{l\ne n}\frac{(m-m_l)}{(m_n-m_l)}.
\end{eqnarray}
To obtain the free energy we also need to determine $\pi$ from Eq. \ref{qmu}. This equation depends on $\mathcal{Z}_{i\rightarrow a}(m)$ which again can be obtained in the above interpolation approximation.
Using the above approximation we can obtain $\pi$ and the free energy as long as $0 \le m \le m_f$. Here $m_f$ is the maximum value of $m$ such that frozen variables do exist. Indeed for $m>m_f$ the fraction of frozen variables is zero and with a trivial $\rho(r)$ we find a zero complexity which is not always correct.
The number of interpolation points is chosen such that the resulted complexity has a reasonable behavior.
In Fig. \ref{figure6} we compare the complexity obtained in this way with the one we obtained by the population dynamics. As the figure shows with the interpolation approximation we are able to reproduce the population dynamics results in the interval $0\le m\le m_f$.
With the above approximation we can find an estimate of the freezing point, $L_f$, where all clusters become frozen. In table \ref{table2} we have compared $L_f$ with degree values at the other phase transition points.

\begin{table}
\begin{center}

\begin{tabular}{|c|c|c|c|c|c|}
  \hline

      $K$  & $L_d$ & $L_r$ & $L_f$ & $L_c$ & $L_s$   \\
  \hline

   $3$   & $6$ & $-$ & $-$ & $7$ & $7$   \\

  \hline

   $4$    & $18$ & $-$ & $-$ & $20$ & $20$   \\

  \hline

   $5$    & $49$ & $53$ & $-$ & $53$ & $53$   \\

  \hline

   $6$    & $114$ & $119$ & $126$ & $130$ & $130$   \\

  \hline

   $7$    & $250$ & $257$ & $297$ & $306$ & $307$  \\

  \hline

   $8$    & $534$ & $543$ & $663$ & $705$ & $706$  \\

  \hline

   $9$    & $1122$ & $1136$ & $1473$ & $1591$ & $1592$  \\

  \hline

  $10$    & $2333$ & $2356$ & $3202$ & $3543$ & $3543$  \\

  \hline

\end{tabular}

\vskip 0.5cm

\caption{Numerical values of degree $L$ at different transition points obtained in the 1RSB approximation with the methods described in the manuscript.}\label{table2}
\end{center}
\end{table}

\section{Algorithms and the entropy landscape}\label{5}

In this section we will use different algorithms to find some solutions of the bicoloring problem close to the SAT/UNSAT transition. We show that a smoothed BP decimation algorithm is able to find solutions even beyond the rigidity transition $L>L_r$. We will also see that, within our level of approximation and for  fixed parameters, the algorithm always finds solutions that belong to the same kind of clusters. Interestingly enough, beyond  the rigidity transition, we find solutions to clusters that are exponentially smaller in number compared to the thermodynamically relevant ones.

\subsection{Cavity method as an algorithm to find solutions}\label{51}

\begin{figure}
\includegraphics[width=10cm]{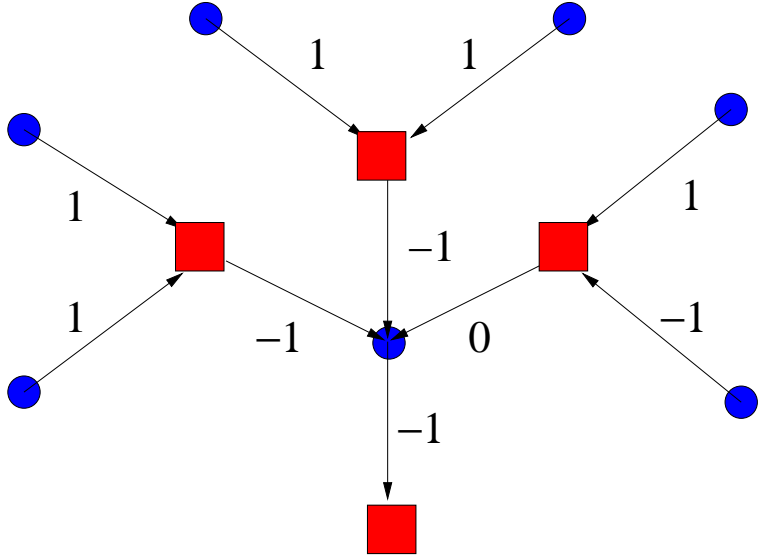}
\caption{Warning propagation on a factor graph.}\label{figure7}
\end{figure}

\textit{Warning Propagation} (WP) is an elementary message passing algorithm that uses cavity messages to find a solution of a constraint satisfaction problem. On each edges of the factor graph we define cavity messages $W_{a\rightarrow i}\in \{-1,0,1\}, W_{i\rightarrow a} \in \{-1,1\}$; The warning $W_{a\rightarrow i}=0$ means that variable $i$ is free to take any value without worrying about constraint $a$. On the other hand, if $W_{a\rightarrow i}=-1,1$, variable $i$ should take a value that satisfies constraint $a$. The message $W_{i\rightarrow a}=-1,1$ represents the color that variable $i$ has to take to satisfy the other constraints.  Given a factor graph we start with the initially random values of $W$'s and in each sweep of the algorithm we update all the messages, see Fig. \ref{figure7}. For example the messages on edge $(i,a)$ are updated in the following way:
\begin{eqnarray}
W_{a\rightarrow i}=\left\{ \begin{array}{ll}
$-1$ & \hbox{if all $W_{j\rightarrow a}$'s are $1$},\\
$1$ & \hbox{if all $W_{j\rightarrow a}$'s are $-1$},\\
$0$ & \hbox{otherwise},
\end{array}  \right.  \\ \nonumber
W_{i\rightarrow a}
=\it{sign}(\sum_{b\in V(i) \setminus a} W_{b\rightarrow i}).
\end{eqnarray}

If the algorithm converges and no variable receives contradictory warnings we can determine the solution according to the warnings.
It has been shown that on tree factor graphs the above algorithm always converges and gives the solutions.

More sophisticated message passing algorithms that work much better than WP are \textit{Belief Propagation Decimation} (BPD) and \textit{Survey Propagation Decimation} (SPD) \cite{BMZ-rsalgorithm-2005}. In these algorithms one replaces the messages $W_{a\rightarrow i}, W_{i\rightarrow a}$ with probabilities that come from single cluster (RS) or multiple cluster (1RSB) approximation. For example, in BPD we have the believes $\mu_{a\rightarrow i}, \mu_{i\rightarrow a}$ that are updated according to the BP equations
\begin{eqnarray}\label{BP}
\mu_{i\rightarrow a}(\sigma_i)=\frac{1}{Z_{i\rightarrow
a}}\prod_{b\in V(i)\setminus a} \mu_{b\rightarrow i}(\sigma_i), \\ \nonumber
\mu_{b\rightarrow i}(\sigma_i)=\sum_{\sigma_{\partial b \setminus i}}
I_b(\sigma_{\partial b})\prod_{j\in V(b)\setminus i} \mu_{j\rightarrow
b}(\sigma_j).
\end{eqnarray}
Starting from random initial values for the $\mu$'s we update them to reach a fixed point of the dynamics.  After convergence we define the local fields
\begin{eqnarray}
H_i\equiv \ln\frac{\mu_i(1)}{\mu_i(0)},
\end{eqnarray}
where
\begin{eqnarray}
\mu_{i}(\sigma_i)=\frac{1}{Z_i}\prod_{b\in V(i)} \mu_{b\rightarrow i}(\sigma_i).
\end{eqnarray}
Then the most biased variable is fixed according to the sign of its local field. Then we simplify the factor graph and repeat the above procedure till we obtain a paramagnet where $H_i=0$ for all the remained variables. At this stage we can run a local search algorithm to complete the solution of our problem.
In this paper we are going to use a smoothed version of BPD first introduced in \cite{BZ-prl-2006}. The main idea is to introduce external fields $h_i$ that at each step of the algorithm are updated according to the local fields. At the end, the external fields determine the preferred values of the variables.
We call this algorithm  \textit{Belief Propagation Reinforcement} \cite{BKMZ-isit-2007}.

More precisely,  the BP Reinforcement algorithm works as follows:

\begin{itemize}

\item Start with random initial values for $0\le \mu_{i\rightarrow a},\mu_{a\rightarrow i} \le 1$ and $-\delta \le h_i \le \delta$ ($\delta \ll 1$).

\item For $t=1,\ldots,t_{max}$ :

\begin{itemize}

\item Update all the $\mu$'s according to the BP equations in presence of the external fields:
\begin{eqnarray}\label{BPh}
\mu_{i\rightarrow a}(\sigma_i)=\frac{e^{h_i \sigma_i}}{Z_{i\rightarrow
a}}\prod_{b\in V(i)\setminus a} \mu_{b\rightarrow i}(\sigma_i), \\ \nonumber
\mu_{b\rightarrow i}(\sigma_i)=\sum_{\sigma_{\partial b\setminus i}}
I_b(\sigma_{\partial b})\prod_{j\in V(b)\setminus i} \mu_{j\rightarrow
b}(\sigma_j).
\end{eqnarray}

\item Obtain local fields $H_i=\ln\frac{\mu_i(1)}{\mu_i(0)}$ and with probability $1-t^{-\gamma}$ update the external fields as $h_i\rightarrow h_i+\it{sign}(H_i) \delta$.
\item If $\underline{\sigma}=\{\sigma_i=\it{sign}(H_i)|i=1,\ldots,N\}$ is a solution, return SOLUTION $=\underline{\sigma}$.
\end{itemize}

\end{itemize}
Notice that instead of fixing variables one by one, here all the external fields are updated (with a rate that increases with $t$) during the run time. Moreover, in this algorithm we do not need to simplify the factor graph after each decimation.

For comparison, we also use other algorithms like \textit{Simulated Annealing} (SA) and \textit{Focused Simulated Annealing} (FSA) to find solutions \cite{SAO-jstat-2005}.
In the SA algorithm we start from a random configuration at small inverse temperature $\beta_i\equiv 1/T_i$ and decrease the temperature slowly. At each temperature we select all the variables in a random sequential way and flip a variable with probability $\textit{min}\{1,\exp(-\beta \Delta E_i)\}$. Here $\Delta E_i$ is the change in the number of unsatisfied constraints, if we accept to flip variable $i$. After a sweep the inverse temperature increases by $\Delta \beta$.
In the FSA algorithm we do the same as SA except that to flip a variable we only select those that belong to unsatisfied constraints.

To check the algorithms and their solutions we consider two different cases; (a) $(4,19)$-hypergraphs ($L_d<L<L_s$), just after the dynamical transition and before the SAT/UNSAT transition where the thermodynamically relevant clusters are unfrozen. (b) $(6,121)$-hypergraphs ($L_r<L<L_f$), where the thermodynamically relevant clusters are frozen but still there are some unfrozen clusters.

In case (a) we are able to find some solutions with all BPR, SA and FSA algorithms in a reasonable time. On a $(4,19)$-hypergraph of $N=10^4$ variables it takes about $20$ hours for SA and FSA algorithms to find a solution whereas BPR algorithm does the job in about $10$ minutes. In SA and FSA algorithms the parameters are $\beta_i=0.1$ and $\Delta \beta=10^{-5}$. In BPR algorithm we used $\gamma=0.05$ and $\delta=0.01$. With these parameters we could obtain a solution at the end of almost all runs.

In case (b) we could only find solutions with the BPR algorithm . Both SA and FSA algorithms were not able to give a solution in a couple of days even for $N=10^3$. However, on a hypergraph of size $N=10002$, the BPR algorithm still returns a solution for every instance after a few hours in about $20$ percent of the runs starting with different initial conditions.

We expect the performance of the algorithm could be improved by further optimization. We did not pursue this line as we are interested in a proof of concept rather than in optimizing algorithms over academic benchmarks.

\subsection{Entropy versus distance from a solution}\label{52}

\begin{figure}
\includegraphics[width=10cm]{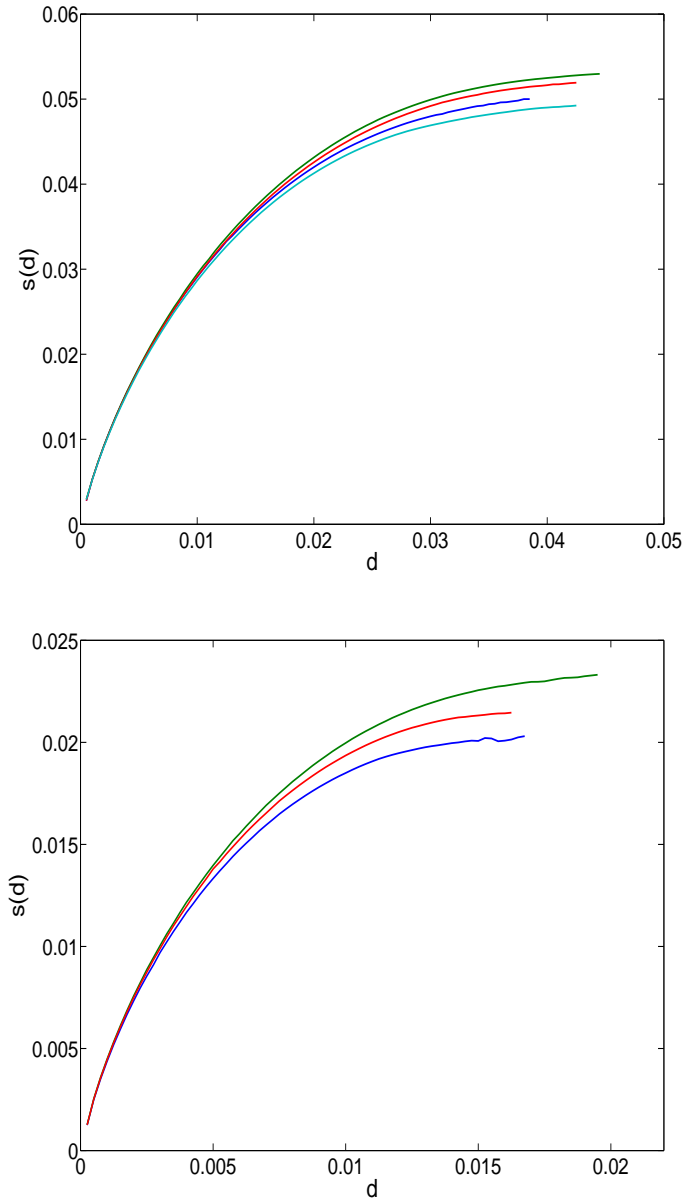}
\caption{$s(d)$ for a few solutions in a $(4,19)$-hypergraph with $N=10000$ (top) and a $(6,121)$-hypergraph with $N=10002$ (bottom). Please note that some of the curves have been selected to show the extremal behavior of $s(d)$. }\label{figure8}
\end{figure}

Suppose that we have the number of solutions at distance $d$ from a given solution, $e^{Ns(d)}$. If the solution belongs to a sphere-like cluster of solutions then $s(d)$ increases for $d\le d^*$ and becomes zero at $d^*$. Clearly, for large $N$, the entropy $s^*$ is a good (under)estimate of the total entropy of the cluster. If the solution space is more complex we still expect $s(d)$ to be, up to a distance $d^*$, an increasing function of $d$. It may exhibit a maximum at $d^*$ and decrease for larger distances. In any case we can take $s^*$ as an approximation to the entropy of the cluster.

To obtain $s(d)$ for a given solution $\underline{\sigma}^*$ and distance $d$, we use a local BP as follows:

\begin{itemize}

\item Start with random initial values for $0\le \mu_{i\rightarrow a},\mu_{a\rightarrow i} \le 1$ and a reasonable value of $x$.

\item For $t=1,\ldots,t_{max}$ :

\begin{itemize}

\item Update all the $\mu$'s according to the BP equations around the given solution:
\begin{eqnarray}
\mu_{i\rightarrow a}(\sigma_i)=\frac{e^{x(\sigma_i-\sigma_i^*)^2}}{Z_{i\rightarrow
a}}\prod_{b\in V(i)\setminus a} \mu_{b\rightarrow i}(\sigma_i),\\ \nonumber
\mu_{b\rightarrow i}(\sigma_i)=\sum_{\sigma_{\partial b\setminus i}}
I_b(\sigma_{\partial b})\prod_{j\in V(b)\setminus i} \mu_{j\rightarrow
b}(\sigma_j).
\end{eqnarray}

\item Obtain $f(x)$ (Eq. \ref{free_energy1}) and find the new $x$ such that $d=\frac{\partial f(x)}{\partial x}$.

\item If converged, calculate the entropy (Eq. \ref{entropy1}) and return $s(d)$.

\end{itemize}

\end{itemize}

The SP version of this algorithm has been used in \cite{MM-jstatmech-2006} to obtain the complexity as a function of distance from a solution in $k$-XOR-SAT problem.

Figure \ref{figure8} shows $s(d)$ for a number of solutions obtained with BPR algorithm.  As the figures show, we do not always reach the extrema point of the curve $s(d)$. It is even more difficult to observe the decreasing part of the entropy. Indeed, as we approach the maximum, the convergence time of the algorithm increases rapidly and exceeds the upper bound $t_{max}=1000$. This happens, probably, when we encounter the other clusters where replica symmetry approximation is not valid any more.  However, we could observe the decreasing part of $s(d)$ for small values of $N$, where computational time is not too large.

Finally, notice that one could obtain the cluster entropy by summing over all solutions: $\exp(Ns)=\sum_d \exp(Ns(d))$. However, for large $N$ the maximum entropy has the dominant contribution to the cluster entropy. To show this we calculated the cluster entropy $s$, using the above definition, and compared it with our estimation $s^*$ which is the maximum entropy. For instance, for three solutions of a $(4,19)$-hypergraph of size $N=10000$ we obtain $\delta s=s-s^*=0.000115, 0.000118, 0.000121$ whereas $s^*=0.052, 0.050, 0.0497$, respectively. We see that the differences are very small compared to the cluster entropy. 

Another source of systematic error is that the curves do not always reach the real extremum. However, as figure \ref{figure8} shows, we observed that the maximum entropy is very close to the real one. Indeed an extrapolation of the curves to higher distances gives a correction which is about $0.001$.

\subsection{$m=1$ vs $m \neq 1$ solutions}\label{53}

\begin{figure}
\includegraphics[width=15cm]{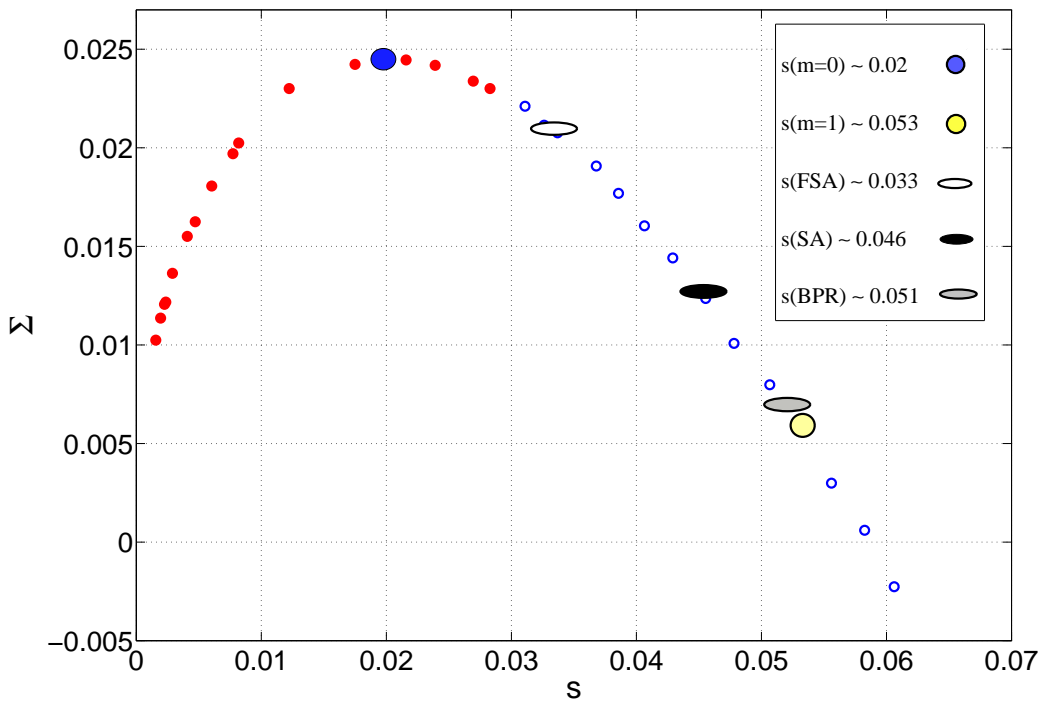}
\caption{Comparing the attractive clusters of different algorithms in the entropy landscape. The large circles show the typical and thermodynamically relevant clusters. All FSA, SA and BPR algorithms find solutions in the interval between frozen and thermodynamically relevant clusters with entropies $s(FSA)<s(SA)<s(BPR)$. In each case we found $20$ solutions on a $(4,19)$-hypergraph of size $N=10000$. The standard deviations in the entropies is about $0.002$}\label{figure9}
\end{figure}

Using the method described in the previous subsection we can now locate our solutions in the entropy landscape to see to which clusters they belong. In Fig. \ref{figure9} we show the attractive clusters of different algorithms after the dynamical transition and before the rigidity transition. In this case BPR finds solutions in clusters that are very close to the thermodynamically relevant ones. We think that the difference is due to the systematic errors in underestimating the cluster entropy. 
In addition, there is also some statistical error in the entropy value of the curve points.
The figure also shows that SA ends up in smaller clusters compared to the thermodynamically relevant ones. Moreover, FSA finds solutions in much smaller clusters close to the frozen ones.

The above results have been obtained with parameters given in section \ref{51}. We found that by decreasing $\gamma$ (in BPR) or $\Delta \beta$ (in SA and FSA) the algorithms find solutions in larger clusters. In fact in a very slow annealing scheme, where one equilibrates the system at each step of the algorithm, we will finally find a solution in the thermodynamically relevant clusters.

Notice that all the algorithms end up in the region between the frozen and thermodynamically relevant clusters. Indeed, when $N$ is large it is very difficult to find a solution in the frozen clusters; each time we flip a frozen variable we go to another cluster of solutions and so an extensive number of flips is needed to accordingly rearrange the variables. On the other hand, it is not also easy to find a solution in very large clusters that are exponentially less numerous than the thermodynamically relevant ones.

\begin{figure}
\includegraphics[width=15cm]{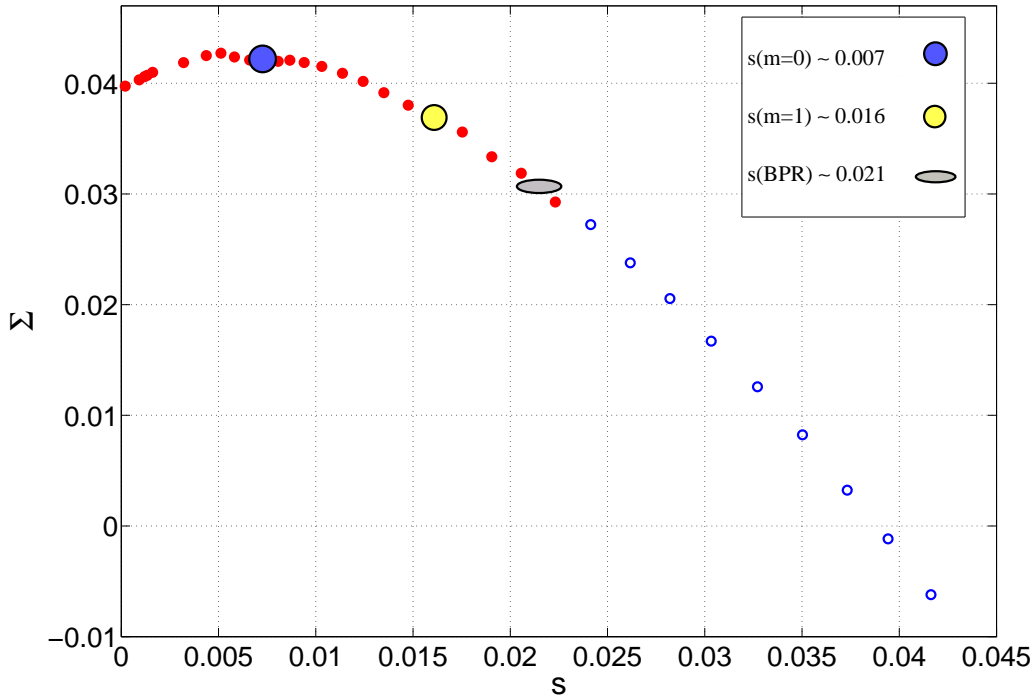}
\caption{Comparing the attractive clusters of the BPR algorithm with the typical and thermodynamically relevant clusters (large circles). In this case the BPR algorithm finds solutions in the most numerous unfrozen clusters. The result obtained from $20$ solutions on a $(6,121)$-hypergraph of size $N=10002$. The standard deviation in the entropy is about $0.002$.}\label{figure10}
\end{figure}

As already mentioned, beyond the rigidity transition we could only find a solution by BPR algorithm. Figure \ref{figure10} shows that in this case the solutions are very close to the boundary between frozen and unfrozen clusters. The difference is about the statistical errors and the error that we make by underestimating the entropy.  We see that when the thermodynamically relevant clusters are frozen the algorithm ends up in the smallest unfrozen clusters. These are exponentially more numerous than the other unfrozen clusters. 

We have checked that our solutions do, indeed, belong to the unfrozen clusters. This can be done with the so called {\em Whitening}  process \cite{P-arxiv-2002,AR-proceeding-2006}. Given a solution one checks if a variable can be flipped without violating any constraint. If so, that variable is unfrozen and is denoted with '$*$'. The process goes on by checking one by one the other variables with the additional rule that a constraint with at least one star variable is always satisfied. This process is repeated up to the fixed point where the number of star variables is fixed.
If a solution belongs to an unfrozen cluster then at the end all the variables would be '$*$'.

\section{Conclusion}\label{6}

In summary we applied the large deviations cavity method to study the phase diagram of the bicoloring problem on regular random hypergraphs.
Working in the one-step replica symmetry breaking framework we located the various phase transitions characterizing the structure of the solutions landscape at both the ensemble and single instance level. Notice that we did not check the stability of 1RSB solutions toward higher order replica symmetry breaking. But, as other studies show \cite{MR-epjb-2003,MPR-jphysa-2004}, we expect 1RSB solutions to give the correct qualitative picture and even exact results close to the SAT/UNSAT transition.

We also used different algorithms to find solutions and to locate them in the entropy landscape of the problem. This provided a rough characterization of the relations existing between the entropic properties of clusters of solutions and the different algorithms used to find them.

From an algorithmic point of view, the algorithms based on simulated annealing could not efficiently find solutions after the rigidity point. However, using BPR  we showed that it is actually possible to go beyond the rigidity transition. In this case we obtained  solutions that belong to the smallest and most numerous unfrozen clusters \cite{S-arxiv-2007}.

\acknowledgments

We thank T. Mora and L. Zdeborova that let us know about their numerical methods.
The work was supported by the EVERGROW EC grant and by the Technical Computing Initiative of Microsoft Research.

\renewcommand{\theequation}{A-\arabic{equation}}
\setcounter{equation}{0}
\appendix

\section{Cavity equations in the RS approximations}\label{app1}

We start from the partition function definition in Eq. \ref{Z} and derive the main equations in the first part of section \ref{2}.
Let $Z_{i\rightarrow a}(\sigma_i)$ denote the partition function in the absence of constraint $a$ and when variable $i$ has state $\sigma_i$. Then, in the absence of constraint $a$, the probability of finding  variable $i$ in state $\sigma_i$ is
\begin{equation}
\mu_{i\rightarrow a}(\sigma_i)=\frac{Z_{i\rightarrow
a}(\sigma_i)}{\sum_{\sigma}Z_{i\rightarrow a}(\sigma)}.
\end{equation}

On the other hand, assuming a tree structure for the factor graph we can write
\begin{equation}\label{app-Zia}
Z_{i\rightarrow a}(\sigma_i)=\sum_{\sigma_{\partial i\rightarrow a}}\left(\prod_{b\in
V(i)\setminus a} I_b(\sigma_{\partial b})[\prod_{j\in V(b)\setminus i}
Z_{j\rightarrow b}(\sigma_j)]\right)e^{x(\sigma_i-\sigma_i^*)^2}.
\end{equation}

From the above equation we can derive a relation for the cavity marginals 

\begin{equation}\label{app-muia}
\mu_{i\rightarrow a}(\sigma_i)=\frac{1}{Z_{i\rightarrow
a}}\sum_{\sigma_{\partial i\rightarrow a}}\left(\prod_{b\in V(i)\setminus a}
I_b(\sigma_{\partial b })[\prod_{j\in V(b)\setminus i} \mu_{j\rightarrow
b}(\sigma_j)]\right)e^{x(\sigma_i-\sigma_i^*)^2},
\end{equation}

where $Z_{i\rightarrow a}$ is a normalization constant. 
It is more convenient if we write the above relation as

\begin{equation}\label{app-mu1}
\mu_{i\rightarrow a}(\sigma_i)=\frac{e^{x(\sigma_i-\sigma_i^*)^2}}{Z_{i\rightarrow
a}}\prod_{b\in V(i)\setminus a} \mu_{b\rightarrow i}(\sigma_i),
\end{equation}

where

\begin{equation}\label{app-mu2}
\mu_{b\rightarrow i}(\sigma_i)=\sum_{\sigma_{\partial b\setminus i}}
I_b(\sigma_{\partial b})\prod_{j\in V(b)\setminus i} \mu_{j\rightarrow
b}(\sigma_j).
\end{equation}

The free energy  $f(x)$ is given by $\frac{1}{N}\ln Z$. In the Bethe approximation

\begin{equation}\label{app-bethef}
f(x)=\sum_i \Delta f_i-\sum_a (K_a-1)\Delta f_a,
\end{equation}

where $\Delta f_i$ and $\Delta f_a$ are the free energy shifts by adding variable node $i$ and function node $a$, respectively.

Suppose that we have removed node $i$ and all its function nodes from the factor graph. In this case the partition function reads

\begin{equation}
Z_{-\{i,V(i)\}}=\prod_{a \in V(i)}[\prod_{j\in
V(a)\setminus i}Z_{j\rightarrow a}],
\end{equation}

whereas in the complete factor graph

\begin{equation}
Z=\sum_{\sigma_i,\sigma_{\partial i}}\prod_{a \in
V(i)}[I_a(\sigma_{\partial a})\prod_{j\in V(a)\setminus i}Z_{j\rightarrow
a}(\sigma_j)]e^{x(\sigma_i-\sigma_i^*)^2}.
\end{equation}

Dividing the two quantities we get

\begin{equation}
\frac{Z}{Z_{-\{i,V(i)\}}}=e^{N\Delta
f_i}=\sum_{\sigma_i,\sigma_{\partial i}}\prod_{a \in
V(i)}[I_a(\sigma_{\partial a})\prod_{j\in
V(a)\setminus i}\frac{Z_{j\rightarrow a}(\sigma_j)}{Z_{j\rightarrow
a}}]e^{x(\sigma_i-\sigma_i^*)^2},
\end{equation}

and this gives the shift in the free energy by adding  variable node $i$

\begin{equation}\label{app-dfi}
e^{N\Delta f_i}=\sum_{\sigma_i}\prod_{a \in V(i)}\mu_{a\rightarrow
i }(\sigma_i)e^{x(\sigma_i-\sigma_i^*)^2}.
\end{equation}

We can do the same procedure for a function node. If we remove function node $a$ from the factor graph we have

\begin{equation}
Z_{-a}=\prod_{i \in V(a)}Z_{i\rightarrow a},
\end{equation}

whereas the complete partition function can be written as

\begin{equation}
Z=\sum_{\sigma_{\partial a}}I_a(\sigma_{\partial a})\prod_{i \in
V(a)}Z_{i\rightarrow a}(\sigma_i).
\end{equation}

So

\begin{equation}
\frac{Z}{Z_{-a}}=e^{N\Delta f_a}=\sum_{\sigma_{\partial
a}}I_a(\sigma_{\partial a})\prod_{i \in V(a)}\frac{Z_{i\rightarrow
a}(\sigma_i)}{Z_{i\rightarrow a}},
\end{equation}

and the shift in the free energy is given by

\begin{equation}\label{app-dfa}
e^{N\Delta f_a}=\sum_{\sigma_{\partial a}}I_a(\sigma_{\partial
a})\prod_{i \in V(a)}\mu_{i\rightarrow a}(\sigma_i).
\end{equation}

Finally using Eqs. \ref{app-bethef}, \ref{app-dfi} and \ref{app-dfa} we obtain the free energy $f(x)$ in the Bethe approximation

\begin{eqnarray}\label{app-free_energy1}
Nf(x)=\sum_i \ln Z_i-\sum_a
(K_a-1)\ln Z_a, \\ \nonumber
Z_i\equiv \sum_{\sigma_i}[\prod_{a\in V(i)}
\mu_{a\rightarrow
i}(\sigma_i)]e^{x(\sigma_i-\sigma_i^*)^2},\\ \nonumber
Z_a\equiv \sum_{\sigma_{\partial a}}I_a(\sigma_{\partial
a})\prod_{i\in V(a)} \mu_{i\rightarrow a}(\sigma_i).
\end{eqnarray}

\renewcommand{\theequation}{B-\arabic{equation}}
\section{Cavity equations in the 1RSB approximation}\label{app2}

We start from the generalized partition function in Eq. \ref{mathcalZ} and   
explain the main equations in the second part of section \ref{2}. 

In the Bethe approximation

\begin{equation}
N\mathcal{F}(m)=\sum_i \Delta \mathcal{F}_i-\sum_a (K_a-1)\Delta
\mathcal{F}_a.
\end{equation}

The generalized partition function can be written as
\begin{equation}
\mathcal{Z}=\sum_{c} e^{m N[s_{c,-\{i,V(i)\}}+\Delta
s_{c,i}]},
\end{equation}

where $\Delta s_{c,i}$ is the shift in the entropy of cluster $c$ by adding node $i$ and all its function nodes.
At a fixed value of $m$ the typical clusters would have nearly the same entropy and it seems safe to approximate $e^{m N\Delta s_i}$ with its average value among the clusters, i.e.

\begin{equation}
\mathcal{Z}=[\sum_{c} e^{m Ns_{c,-\{i,V(i)\}}}]\int
d\mathcal{P}[\mu] e^{m N\Delta s_i},
\end{equation}
where $\mathcal{P}[\mu]$ is the probability distribution of $\mu_{i\rightarrow a}$'s among the clusters.
So the shift in the generalized free energy reads
\begin{equation}\label{app-dffi}
e^{N\Delta \mathcal{F}_i}=\int \prod_{a\in V(i)}\prod_{j\in
V(a)\setminus i} d\mathcal{P}_{j\rightarrow a}[\mu_{j\rightarrow a}]
e^{mN\Delta s_i}.
\end{equation}

In the case of adding a function node, similarly we find
\begin{equation}\label{app-dffa}
e^{N\Delta \mathcal{F}_a}=\int \prod_{i \in
V(a)}d\mathcal{P}_{i\rightarrow a}[\mu_{i\rightarrow a}]
e^{mN\Delta s_a}.
\end{equation}

Notice that $\Delta s_i$ and $\Delta s_a$ correspond to the free energy shifts, Eqs. \ref{app-dfi} and \ref{app-dfa}, with $x=0$.
Using Eqs. \ref{app-dffi} and \ref{app-dffa} along with the Bethe form of the generalized free energy we obtain

\begin{eqnarray}\label{app-free_energy2}
N\mathcal{F}(m)=\sum_i \ln \mathcal{Z}_i-\sum_a (K_a-1)\ln \mathcal{Z}_a,\\ \nonumber
\mathcal{Z}_i\equiv \int D_i\mathcal{P}[\mu] e^{mN\Delta
s_i}, \\ \nonumber
\mathcal{Z}_a\equiv \int D_a\mathcal{P}[\mu] e^{mN\Delta s_a},
\end{eqnarray}

where $d\mathcal{P}[\mu]$ is determined by Eq. \ref{pmu}. The normalization constant in this equation is
\begin{equation}\label{app-mathcalZia}
\mathcal{Z}_{i\rightarrow a}=\int \prod_{b\in
V(i)\setminus a}\prod_{j\in V(b)\setminus i} d\mathcal{P}_{j\rightarrow
b}[\mu] e^{mN\Delta s_i}.
\end{equation}
We represented  $\mathcal{P}[\mu]$ as
\begin{equation}
\mathcal{P}[\mu]=\frac{1-\pi}{2}[\delta(r)+\delta(r-1)]+\pi \rho(r),
\end{equation}
where $\rho(r)$ is the probability distribution of unfrozen marginals. By the normalization and symmetry of the problem
\begin{eqnarray}
\int dr  \rho(r) =1,\\ \nonumber
\int dr  r\rho(r) =\int dr (1-r)\rho(r) =\frac{1}{2}.
\end{eqnarray}
Then for the $\mathcal{Q}_{i\rightarrow a}^{\sigma}[\mu]=2\mu_{i\rightarrow a}(\sigma)\mathcal{P}_{i\rightarrow a}[\mu]$ we  have

\begin{eqnarray}
\mathcal{Q}_{i\rightarrow a}^{0}[\mu]=(1-\pi)\delta(1-r)+2\pi r \rho(r),\\ \nonumber
\mathcal{Q}_{i\rightarrow a}^{1}[\mu]=(1-\pi)\delta(r)+2\pi (1-r) \rho(r).
\end{eqnarray}
Notice that for $\sigma \in \{0,1\}$
\begin{equation}
\int d\mu \mathcal{Q}_{i\rightarrow a}^{\sigma}[\mu]=1.
\end{equation}

\renewcommand{\theequation}{C-\arabic{equation}}
\section{Calculating the generalized free energy}\label{app3}

To calculate the free energy $\mathcal{F}$ we need to obtain $\mathcal{Z}_i$, $\mathcal{Z}_a$, the fraction of unfrozen variables $\pi$ and $\rho(r)$.  Let us start from Eq. \ref{pmu} and multiply both sides of the equation with $e^{-rt}$. Integrating over $r$ allows us to get rid of delta function and we find

\begin{eqnarray}\label{app-pmu}
\frac{1-\pi}{2}(1+e^{-t})+\pi \int \rho(r) e^{-rt}dr \hskip 3cm \\
\nonumber  =\frac{1}{\mathcal{Z}_{i\rightarrow a}}\prod_{b\in V(i)\setminus a}
[\sum_{n_b^0,n_b^1,n_b^*}  \left(\begin{array}{c}
K-1\\
n_b^0,n_b^1,n_b^*
\end{array}\right)  (\frac{1-\pi}{2})^{n_b^0+n_b^1}\pi^{n_b^*} 
\prod_{j \in V^*(b)} \int dr_b^j \rho(r_b^j)]  \\ \nonumber  \times [\prod_b \lambda_0(n_b,r_b)+\prod_b \lambda_1(n_b,r_b)]^me^{-\frac{t}{Z_{i\rightarrow a}} \prod_b \lambda_0(n_b,r_b)} ,
\end{eqnarray}
where $n_b^0$ and $n_b^1$ are the number of frozen variables in $V(b)\setminus i$ that take values $0$ and $1$, respectively. Accordingly $V^*(b)\setminus i$ is the set of unfrozen variables in $V(b)\setminus i$ and $n_b^*$ is the number of its elements. Notice that $n_b$'s should satisfy $n^0+n_b^1+n_b^*=K-1$.

Using Eqs. \ref{app-muia} and \ref{app-dfi} we write
\begin{eqnarray}
r=\frac{1}{Z_{i\rightarrow a}} \prod_{b\in V(i)\setminus a} \lambda_0(n_b,r_b),\\ \nonumber
e^{N\Delta s_i}=\prod_{b\in V(i)} \lambda_0(n_b,r_b)+\prod_{b\in V(i)} \lambda_1(n_b,r_b),
\end{eqnarray}

where

\begin{eqnarray}\label{app-lambda01}
\lambda_0(n_b,r_b)=(1-\delta_{n_b^0,K-1})[1-\delta_{n_b^1,0}(1-\delta_{n_b^*,0}) \prod_{j\in V^*(b)\setminus i} r_b^j], \\ \nonumber
\lambda_1(n_b,r_b)=(1-\delta_{n_b^1,K-1})[1-\delta_{n_b^0,0}(1-\delta_{n_b^*,0}) \prod_{j\in V^*(b)\setminus i} (1-r_b^j)].
\end{eqnarray}

One can use Eq. \ref{app-pmu} to write some equations for different moments of $\rho(r)$. For example, for the second moment we obtain
\begin{eqnarray}\label{app-smr}
\frac{1-\pi}{2}+\pi \langle r^2 \rangle  =\frac{1}{\mathcal{Z}_{i\rightarrow a}}\prod_{b\in V(i)\setminus a}
[\sum_{n_b^0,n_b^1,n_b^*} \left(\begin{array}{c}
K-1\\
n_b^0,n_b^1,n_b^*
\end{array}\right) (\frac{1-\pi}{2})^{n_b^0+n_b^1}\pi^{n_b^*} \\ \nonumber \times \prod_{j \in V^*(b)\setminus i} \int dr_b^j \rho(r_b^j)][\prod_{b\in V(i)\setminus a} \lambda_0(n_b,r_b)+\prod_{b\in V(i)\setminus a} \lambda_1(n_b,r_b)]^m(\frac{\prod_b \lambda_0(n_b,r_b) }{Z_{i\rightarrow a}})^2 .
\end{eqnarray}

To compute $\mathcal{F}$ we also need to find $e^{N\Delta s_a}$ in Eq. \ref{app-dfa} 
\begin{equation}
e^{N\Delta s_a}=(1-\delta_{n_a^0,K})(1-\delta_{n_a^1,K})[1-\delta_{n_a^1,0} (1-\delta_{n_a^*,0})\prod_{i\in V^*(a)} r_a^j-\delta_{n_a^0,0} (1-\delta_{n_a^*,0})\prod_{i\in V^*(a)} (1-r_a^i)].
\end{equation}

The normalization constants in Eqs. \ref{app-Zia} and \ref{app-mathcalZia} are
\begin{equation}\label{app-Zia1}
Z_{i\rightarrow a}=\prod_{b\in V(i)\setminus a} \lambda_0(n_b,r_b)+\prod_{b\in V(i)\setminus a} \lambda_1(n_b,r_b),
\end{equation}

and

\begin{eqnarray}\label{app-mathcalZia1}
\mathcal{Z}_{i\rightarrow a}=\prod_{b\in V(i)\setminus a}
[\sum_{n_b^0,n_b^1,n_b^*} \left(\begin{array}{c}
K-1\\
n_b^0,n_b^1,n_b^*
\end{array}\right) (\frac{1-\pi}{2})^{n_b^0+n_b^1}\pi^{n_b^*} \prod_{j \in V^*(b)\setminus i} \int dr_b^j \rho(r_b^j)] \\ \nonumber \times
[\prod_{b\in V(i)\setminus a} \lambda_0(n_b,r_b)+\prod_{b\in V(i)\setminus a} \lambda_1(n_b,r_b)]^m.
\end{eqnarray}

Finally for the main elements of the generalized free energy we have
\begin{eqnarray}\label{app-mathcalZi}
\mathcal{Z}_{i}=\prod_{b\in V(i)}
[\sum_{n_b^0,n_b^1,n_b^*} \left(\begin{array}{c}
K-1\\
n_b^0,n_b^1,n_b^*
\end{array}\right) (\frac{1-\pi}{2})^{n_b^0+n_b^1}\pi^{n_b^*} \prod_{j \in V^*(b)\setminus i} \int dr_b^j \rho(r_b^j)]
\\ \nonumber \times [\prod_{b\in V(i)\setminus a} \lambda_0(n_b,r_b)+\prod_{b\in V(i)\setminus a} \lambda_1(n_b,r_b)]^m,
\end{eqnarray}

and

\begin{eqnarray}\label{app-mathcalZa}
\mathcal{Z}_a=
\sum_{n_a^0,n_a^1,n_a^*} \left(\begin{array}{c}
K\\
n_a^0,n_a^1,n_a^*
\end{array}\right) (\frac{1-\pi}{2})^{n_a^0+n_a^1}\pi^{n_a^*} \prod_{i \in V^*(a)} \int dr_a^i \rho(r_a^i)
[(1-\delta_{n_a^0,K})(1-\delta_{n_a^1,K})\\ \nonumber \times \left(1-\delta_{n_a^1,0} (1-\delta_{n_a^*,0})\prod_{i\in V^*(a)} r_a^j-\delta_{n_a^0,0} (1-\delta_{n_a^*,0})\prod_{i\in V^*(a)} (1-r_a^i)\right)]^m.
\end{eqnarray}

In the following we will give the details of calculations in two special cases that need more explanation.

\subsection{The case $\pi=1$}\label{app31}

When $\pi=1$ the equation for $\mathcal{Z}_i$, Eq. \ref{app-mathcalZi}, is

\begin{eqnarray}
\mathcal{Z}_i=\prod_{a=1,L}[\prod_{j=1,K-1} \int dr_a^j \rho(r_a^j)][\prod_{a}\left(1-\prod_{j} r_a^j\right)+\prod_{a}\left(1-\prod_{j} (1-r_a^j)\right)]^m.
\end{eqnarray}

For $m=0,1,2$ we obtain

\begin{eqnarray}
\mathcal{Z}_i(m=0)=1, \hskip 3cm \\ \nonumber
\mathcal{Z}_i(m=1)=2[1-\frac{1}{2^{K-1}}]^L, \hskip 3cm \\ \nonumber
\mathcal{Z}_i(m=2)=2[1-\frac{2}{2^{K-1}}+\langle r^2 \rangle^{K-1}]^L+2[1-\frac{2}{2^{K-1}}+(\frac{1}{2}-\langle r^2 \rangle)^{K-1}]^L.
\end{eqnarray}

For $\mathcal{Z}_a$ from Eq. \ref{app-mathcalZa} we find

\begin{eqnarray}
\mathcal{Z}_a=[\prod_{i=1,K} \int dr_a^i \rho(r_a^i)][1-\prod_{i} r_a^i-\prod_{i} (1-r_a^i)]^m.
\end{eqnarray}

Again for $m=0,1,2$ 

\begin{eqnarray}
\mathcal{Z}_a(m=0)=1,\\ \nonumber
\mathcal{Z}_a(m=1)=[1-\frac{2}{2^{K}}],\\ \nonumber
\mathcal{Z}_a(m=2)=[1-\frac{4}{2^{K}}+2\langle r^2 \rangle^K+2(\frac{1}{2}-\langle r^2 \rangle)^K].
\end{eqnarray}

To complete the free energy calculations we need to find $\langle r^2 \rangle$ in $m=2$ clusters.
The second moment of $\rho(r)$ can be obtained from Eq. \ref{app-smr},

\begin{eqnarray}
\langle r^2\rangle=\frac{1}{\mathcal{Z}_{i\rightarrow a}}\prod_{b=1,L-1}[\prod_{j=1,K-1} \int dr_b^j \rho(r_b^j)] \hskip 2cm\\ \nonumber \times[\prod_{b}\left(1-\prod_{j} r_b^j\right)+\prod_{b}\left(1-\prod_{j} (1-r_b^j)\right)]^{m-2} [\prod_{b}\left(1-\prod_{j} r_b^j\right)]^2.
\end{eqnarray}

If $m=2$, the exact equation is

\begin{eqnarray}\label{app-m2r2}
\langle r^2\rangle=\frac{1}{\mathcal{Z}_{i\rightarrow a}(m=2)}[1-\frac{2}{2^{K-1}}+\langle r^2\rangle^{K-1}]^{L-1}.
\end{eqnarray}

Now we can use the Lagrange interpolating polynomial to approximate the free energy by

\begin{eqnarray}
\mathcal{F}(m)=\mathcal{F}(m=0)\frac{(m-1)(m-2)}{2}-\mathcal{F}(m=1)m(m-2)+\mathcal{F}(m=2)\frac{m(m-1)}{2}.
\end{eqnarray}

Since $\mathcal{F}(m=0)=0$ we get

\begin{eqnarray}
\mathcal{F}(m)=-\mathcal{F}(m=1)m(m-2)+\mathcal{F}(m=2)\frac{m(m-1)}{2},\\ \nonumber
s(m)=-2(m-1)\mathcal{F}(m=1)+(m-\frac{1}{2})\mathcal{F}(m=2),\\ \nonumber
\Sigma(m)=m^2[\mathcal{F}(m=1)-\frac{1}{2}\mathcal{F}(m=2)].
\end{eqnarray}

\subsection{The case of integer $m$}\label{app32}

Starting from Eq. \ref{app-mathcalZia} we expand $\mathcal{Z}_{i\rightarrow a}$ for integer $m>0$ to get
\begin{eqnarray}
\mathcal{Z}_{i\rightarrow a}=\sum_{l} \left(\begin{array}{c}
m\\
l
\end{array}\right) \prod_{b\in V(i)\setminus a}
[\sum_{n_b^0,n_b^1,n_b^*} \left(\begin{array}{c}
K-1\\
n_b^0,n_b^1,n_b^*
\end{array}\right) (\frac{1-\pi}{2})^{n_b^0+n_b^1}\pi^{n_b^*} \\ \nonumber \times \prod_{j \in V^*(b)} \int dr_b^j \rho(r_b^j)
\lambda_0^{l}(n_b,r_b)\lambda_1^{m-l}(n_b,r_b)].
\end{eqnarray}

To simplify the results we approximate  $\rho(r)$ by $\delta(r-\frac{1}{2})$.
After some simplifications we obtain

\begin{eqnarray}
\mathcal{Z}_{i\rightarrow a}=\sum_{l} \left(\begin{array}{c}
m\\
l
\end{array}\right)
[1-2(\frac{1+\pi}{2})^{K-1}+\pi^{K-1}+\pi^{K-1}(1-\frac{1}{2^{K-1}})^m \\ \nonumber -\pi^{K-1}(1-\frac{1}{2^{K-1}})^l-\pi^{K-1}(1-\frac{1}{2^{K-1}})^{m-l}\\ \nonumber+\sum_{n} \left(\begin{array}{c}
K-1\\
n
\end{array}\right)(\frac{1-\pi}{2})^{K-1-n}\pi^n\left((1-\frac{1}{2^{n}})^l+(1-\frac{1}{2^{n}})^{m-l}\right)]^{L-1}.
\end{eqnarray}

For $\mathcal{Z}_a$ we use Eq. \ref{app-mathcalZa} and again $\rho(r)=\delta(r-\frac{1}{2})$ to get

\begin{eqnarray}
\mathcal{Z}_a=\pi^K[1-\frac{2}{2^{K}}]^m-2\pi^K[1-\frac{1}{2^{K}}]^m \hskip 2cm \\ \nonumber+2\sum_{n}\left(\begin{array}{c}
K\\
n
\end{array}\right)(\frac{1-\pi}{2})^{K-n}\pi^n[1-\frac{1}{2^{n}}]^m
+[1+\pi^K-2(\frac{1+\pi}{2})^K].
\end{eqnarray}

To obtain the free energy we still need to determine $\pi$. From Eq. \ref{qmu} we have

\begin{eqnarray}\label{app-pigm}
1-\pi=\frac{2^{-(K-1)(L-1)+1}}{\mathcal{Z}_{i\rightarrow a}}\sum_{l=1,L-1} \left(\begin{array}{c}
L-1\\
l
\end{array}\right)[(1-\pi)^{K-1}]^l \hskip 2cm \\ \nonumber \times
[\sum_{\sigma_{\partial b\setminus i}} I_b(\sigma_{\partial b\setminus i}) \int \prod_{j\in V(b)\setminus i}d\mathcal{Q}^{\sigma_j}[\mu] \left(1-\delta_{n_b^1,0}(1-\delta_{n_b^*,0})\prod_{j\in V(b)\setminus i} r_b^j\right)^{m-1}]^{L-1-l}.
\end{eqnarray}

Taking $\rho(r)=\delta(r-\frac{1}{2})$ we obtain an equation for $\pi$ and for general $m$.

\begin{eqnarray}
\frac{1-\pi}{2} \hskip 13cm \\ \nonumber =\frac{1}{\mathcal{Z}_{i\rightarrow a}}\{
[1-(\frac{1+\pi}{2})^{K-1} +\sum_{n} \left(\begin{array}{c}
K-1\\
n
\end{array} \right) (\frac{1-\pi}{2})^{K-1-n}\pi^{n}(1-\frac{1}{2^n})^{m}]^{L-1} \\ \nonumber
-[1-(\frac{1+\pi}{2})^{K-1}-(\frac{1-\pi}{2})^{K-1} +\sum_{n} \left(\begin{array}{c}
K-1\\
n
\end{array}\right) (\frac{1-\pi}{2})^{K-1-n}\pi^{n}(1-\frac{1}{2^n})^{m}]^{L-1}\}.
\end{eqnarray}

\end{document}